# Radiative heat transfer of spherical particles mediated by fluctuation electromagnetic field

G.V.Dedkov and A.A.Kyasov

Nanoscale Physics Group, Kabardino –Balkarian State University, Nalchik,

360004, Russian Federation

We calculate intensity of radiative heat transfer and radiative conductance in a system of two spherical particles embedded in equilibrium vacuum background (photon gas). The temperatures of the particles and of the background radiation are arbitrary. The calculations are based on the dipole and additive approximations of the fluctuation electromagnetic theory. We obtained much higher radiative conductance between $25\,\mu m$ silica particles (by 4 orders of magnitude) in comparison with recent results by A.Narayanaswamy and Gang Chen ( Phys.Rev., 2008).

## 1. Introduction

To date, the problem of radiative heat transfer between condensed bodies at small separations has attracted growing attention both theorists [1-4] and experimentalists [5-7]. Quite recently, the authors [6,7] have reported on first measurements of heat transfer between two silica spheres with diameters of 50 $\mu m$ separated by gaps ranging from 0.1 to 10 $\mu m$ . At such conditions, near –field heat transfer plays dominating role, being mediated by surface phonon –polaritons of the adjacent surfaces. The accuracy and control of numerical calculations of the heating rate $dQ/dt$ are performed by direct comparison with the known analytical solutions in several simple geometries: two half –spaces [8], a small sphere above a plane [4,9] , two spheres in the dipole approximation [10-12]. The second configuration is studied using a proximity approximation [2] similar to the known proximity force approximation [13]. Another approach is based on additive summation method using an exact analytical solution in a point –dipole problem [4,9]. However, the third problem has been solved in [10-12] with no allowance for surrounding uniform vacuum background. Moreover, the obtained expressions for $dQ/dt$ differ by a numerical factor $2\pi$ (compare the results [10] and [11,12]). By the way, an additional heat exchange between the particles and vacuum background essentially changes the long –range asymptotics of the heat transfer and must be taken into account to interpret the results [6,7] in a correct manner.

The aim of this paper is to derive a new expression for heating rates of two spherical particles with radii $R_1, R_2$ and temperatures $T_1, T_2$, embedded in vacuum background filled by an equilibrium photon gas with temperature $T_3$, in dipole approximation. In view of the obtained results, some comments on recent numerical calculations in the two –sphere problem [3] and experimental results [6,7] and are given.

## 2. Theory

The configuration of the system is shown in Fig.1. Both particles are characterized by frequency –dependent electric $\alpha_{ie}(\omega)$ and magnetic $\alpha_{im}(\omega)$ polarizabilities (i=1,2). It is worth noting that restrictions imposed by dipole approximation imply:

$$R_1, R_2 << R, \ R_1, R_2 << \min(\lambda_{T1}, \lambda_{T2}, \lambda_{T3}),$$

where $\lambda_{T1}, \lambda_{T2}, \lambda_{T3}$ are the characteristic wave –lengths of thermal radiation. Relations between $R$ and $\lambda_{T1}, \lambda_{T2}, \lambda_{T3}$ may be arbitrary. We write starting expression for the heating (cooling) rate of the first particle in the form

$$dQ/dt = \dot{Q}_{13}^{vac} + \dot{Q}_{12} \tag{1}$$

where $\dot{Q}_{13}^{vac}$ denotes the rate of heat exchange with vacuum background and $\dot{Q}_{12}$ – that one caused by interaction between the particles. Eq.(1) reflects mutual independence of thermal fluctuations of random electromagnetic fields generated by particles and vacuum background. This enables to calculate different contributions to resulting expression for $dQ/dt$ in an autonomous way [4,9]. So, a more extended expression for $\dot{Q}_{12}$ reads

$$\dot{Q}_{12} = \langle \dot{\mathbf{d}}_1^{in}(t)\mathbf{E}_2^{sp}(\mathbf{r}_1,t)\rangle + \langle \dot{\mathbf{m}}_1^{in}(t)\mathbf{B}_2^{sp}(\mathbf{r}_1,t)\rangle - \langle \dot{\mathbf{d}}_2^{in}(t)\mathbf{E}_1^{sp}(\mathbf{r}_2,t)\rangle - \langle \dot{\mathbf{m}}_2^{in}(t)\mathbf{B}_1^{sp}(\mathbf{r}_2,t)\rangle \quad (2)$$

where $\mathbf{d}_{1,2}(t), \mathbf{m}_{1,2}(t)$ and $\mathbf{E}_{1,2}(\mathbf{r}_{2,1},t), \mathbf{B}_{1,2}(\mathbf{r}_1,t)$ are the dipole electric (magnetic) moments and components of fluctuation electromagnetic field generated by each particle at the location point of another one. The upper indexes "*in*", "*sp*" denote the induced and spontaneous components, the angular brackets –total quantum and statistical averaging. Individual terms in (2) determine parts of the power dissipated in the volume of the particles, corresponding to work of fluctuation electromagnetic field.

To proceed further, the incoming quantities in Eq. (2) are represented by Fourier integrals over frequency $\omega$, with the Fourier transforms $\mathbf{d}_{1,2}^{in}(\omega)$ and $\mathbf{m}_{1,2}^{in}(\omega)$ of the form

$$\mathbf{d}_1^{in}(\omega) = \alpha_{1e}(\omega)\mathbf{E}_2^{sp}(\mathbf{r}_1,\omega), \quad \mathbf{d}_2^{in}(\omega) = \alpha_{2e}(\omega)\mathbf{E}_1^{sp}(\mathbf{r}_2,\omega) \quad (3)$$

$$\mathbf{m}_1^{in}(\omega) = \alpha_{1m}(\omega)\mathbf{B}_2^{sp}(\mathbf{r}_1,\omega), \quad \mathbf{m}_2^{in}(\omega) = \alpha_{2m}(\omega)\mathbf{B}_2^{sp}(\mathbf{r}_2,\omega) \quad (4)$$

With the help of (2)-(4) we obtain

$$\dot{Q}_{12} = \int_{-\infty}^{+\infty}\frac{d\omega}{2\pi}\int_{-\infty}^{+\infty}\frac{d\omega'}{2\pi}\exp[-i(\omega+\omega')t](-i\omega)\{\alpha_{1e}(\omega)\langle \mathbf{E}_2^{sp}(\mathbf{r}_1,\omega)\mathbf{E}_2^{sp}(\mathbf{r}_1,\omega')\rangle + $$
$$+ \alpha_{1m}(\omega)\langle \mathbf{B}_2^{sp}(\mathbf{r}_1,\omega)\mathbf{B}_2^{sp}(\mathbf{r}_1,\omega')\rangle - \alpha_{2e}(\omega)\langle \mathbf{E}_1^{sp}(\mathbf{r}_2,\omega)\mathbf{E}_1^{sp}(\mathbf{r}_2,\omega')\rangle - \quad (5)$$
$$- \alpha_{2m}(\omega)\langle \mathbf{B}_1^{sp}(\mathbf{r}_2,\omega)\mathbf{B}_1^{sp}(\mathbf{r}_2,\omega')\rangle\}$$

The Descartes projections of electromagnetic fields in Eq.(5) are expressed through the components of spontaneous (induced) moments of the particles and components of retarded Green function for free photons [14]

$$E_{1,i}^{sp}(\mathbf{r}_2,\omega) = -\frac{\omega^2}{\hbar c^2}D_{ij}(\omega,\mathbf{R})d_{1,j}^{sp}(\omega), \quad E_{2,i}^{sp}(\mathbf{r}_2,\omega) = -\frac{\omega^2}{\hbar c^2}D_{ij}(\omega,\mathbf{R})d_{2,j}^{sp}(\omega) \quad (6)$$

$$B_{1,i}^{sp}(\mathbf{r}_2,\omega) = -\frac{\omega^2}{\hbar c^2}D_{ij}(\omega,\mathbf{R})m_{1,j}^{sp}(\omega), \quad B_{2,i}^{sp}(\mathbf{r}_2,\omega) = -\frac{\omega^2}{\hbar c^2}D_{ij}(\omega,\mathbf{R})m_{2,j}^{sp}(\omega) \quad (7)$$

$$D_{ij}(\omega,\mathbf{R}) = -\frac{\hbar c^2}{\omega^2}\left\{-\frac{4\pi}{3}\delta(\mathbf{R})\delta_{ij} + \exp(i\omega R/c)\left[\left(\frac{\omega^2}{c^2 R} + \frac{i\omega}{cR^2} - \frac{1}{R^3}\right)(\delta_{ij} - n_i n_j) + 2\left(\frac{1}{R^3} - \frac{i\omega}{cR^2}\right)n_i n_j\right]\right\}, \quad \mathbf{R} = \mathbf{r}_2 - \mathbf{r}_1, \mathbf{n} = \mathbf{R}/|\mathbf{R}| \quad (8)$$

In (6)-(8) we have used common definitions $c, \hbar, k_B$ for the speed of light, Planck's and Boltzmann's constants.

After inserting (6)-(8) in (5) it is obvious, that the field correlators in Eq.(5) are finally expressed through the correlators of the fluctuating electric and magnetic dipole moments [15]

$$\left\langle d_{l,i}^{sp}(\omega)d_{l,k}^{sp}(\omega')\right\rangle = 2\pi\delta_{ik}\delta(\omega+\omega')\hbar\alpha_{le}''(\omega)\coth(\omega\hbar/2k_B T_l), \quad (9)$$

$$\left\langle m_{l,i}^{sp}(\omega)m_{l,k}^{sp}(\omega')\right\rangle = 2\pi\delta_{ik}\delta(\omega+\omega')\hbar\alpha_{lm}''(\omega)\coth(\omega\hbar/2k_B T_l), \quad l=1,2 \quad (10)$$

For example, correlator $\left\langle \mathbf{E}_1^{sp}(\mathbf{r}_2,\omega)\mathbf{E}_1^{sp}(\mathbf{r}_2,\omega')\right\rangle$ takes the form

$$\left\langle \mathbf{E}_1^{sp}(\mathbf{r}_2,\omega)\mathbf{E}_1^{sp}(\mathbf{r}_2,\omega')\right\rangle = 2\pi\delta(\omega+\omega')\hbar\alpha_{1e}''(\omega)\coth(\omega\hbar/2k_B T_1)\left(-\frac{\omega^2}{\hbar c^2}\right)^2 \cdot D_{ik}(\omega,\mathbf{R})D^*_{ik}(\omega,\mathbf{R}) \quad (11)$$

where an upper star sign denotes a complex conjugated Green function, while the twice primed polarizability functions –the corresponding imaginary parts. Other correlators are written analogously. After further elementary calculations we obtain

$$\dot{Q}_{12} = \frac{2\hbar}{\pi R^6} \int_0^\infty d\omega\, \omega [\alpha_{1e}''(\omega)\alpha_{2e}''(\omega) + \alpha_{1m}''(\omega)\alpha_{2m}''(\omega)] \cdot (3 + (\omega R/c)^2 + (\omega R/c)^4) \cdot$$
$$\cdot [\coth(\hbar\omega/2k_B T_2) - \coth(\hbar\omega/2k_B T_1)] \quad (12)$$

Formula (12) agrees with the results obtained in [10-12] with one exception for a numerical coefficient in front of the integral: this coefficient proved to be $1/(2\pi)^2$ in Ref. [10] and $1/(2\pi)^3$ in Refs. [11,12]. Therefore, Eq.(12) manifests a larger value of the heating rate by 1-2 orders of magnitude as compared to [10-12].

The vacuum contribution to the heating rate $\dot{Q}_{13}^{vac}$ has been calculated in our papers [4,16]. A derivation procedure is quite analogous to that one described above. The simplest way to get the final formula is based on usage of energy conservation and Kirchhoff's laws. Then $\dot{Q}_{13}^{vac}$ can be cast in the form

$$\dot{Q}_{13}^{vac} = I(T_3) - I(T_1), \quad (13)$$

where $I(T_1)$ is given by

$$I(T_1) = \frac{2}{3c^3}\left(\langle(\ddot{\mathbf{d}}^{sp}(t))^2\rangle + \langle(\ddot{\mathbf{m}}^{sp}(t))^2\rangle\right) \quad (14)$$

Eq.(14) describes an average intensity of thermal dipole radiation of the particle with temperature $T_1$ in vacuum space caused by spontaneous fluctuating moments. Evidently, this dipolar radiation leads to the particle cooling, whereas the function $I(T_1)$ is determined by the particle absorption spectrum and $T_1$. On the other hand, $I(T_3)$ represents an average intensity of the vacuum radiation with temperature $T_3$, illuminating the particle. According to the Kirchhoff's law, function $I(T_3)$ should be of the same functional form as $I(T_1)$. The only difference between them may be related with different temperature. This allows to calculate only $I(T_1)$, whereas $I(T_3)$ can be obtained replacing $T_1$ by $T_3$ in final expression for $I(T_1)$. Thus, for the average squared dipole electric moment we get

$$\langle (\ddot{\mathbf{d}}^{sp}(t))^2 \rangle = \int_{-\infty}^{\infty} \frac{d\omega}{2\pi} \int_{-\infty}^{\infty} \frac{d\omega'}{2\pi} \langle \ddot{\mathbf{d}}^{sp}(\omega) \ddot{\mathbf{d}}^{sp}(\omega') \rangle \exp[-i(\omega+\omega')t] ,\tag{15}$$

and with account of (10),

$$\langle (\ddot{\mathbf{d}}^{sp}(t))^2 \rangle = \frac{3\hbar}{\pi} \int_0^{\infty} d\omega\, \omega^4 \alpha_e''(\omega) \coth(\hbar\omega/2k_B T_1) \tag{16}$$

In the same manner, a contribution from spontaneous magnetic moment reads

$$\langle (\ddot{\mathbf{m}}^{sp}(t))^2 \rangle = \frac{3\hbar}{\pi} \int_0^{\infty} d\omega\, \omega^4 \alpha_m''(\omega) \coth(\hbar\omega/2k_B T_1) \tag{17}$$

From (14)-(17) we get

$$I(T_1) = \frac{2\hbar}{\pi c^3} \int_0^{\infty} d\omega\, \omega^4 (\alpha_e''(\omega) + \alpha_m''(\omega)) \coth(\hbar\omega/2k_B T_1) \tag{18}$$

The function $I(T_3)$, according to the Kirchhoff's law, reads

$$I(T_3) = \frac{2\hbar}{\pi c^3} \int_0^{\infty} d\omega\, \omega^4 (\alpha_e''(\omega) + \alpha_m''(\omega)) \coth(\hbar\omega/2k_B T_3) \tag{18a}$$

Inserting (18), (18a) into Eq.(13) yields (for the first particle)

$$\dot{Q}_{13}^{vac} = \frac{4\hbar}{\pi c^3} \int_0^{\infty} d\omega\, \omega^4 (\alpha_{1e}''(\omega) + \alpha_{1m}''(\omega)) \left[ \frac{1}{\exp(\hbar\omega/k_B T_3)-1} - \frac{1}{\exp(\hbar\omega/k_B T_1)-1} \right] \tag{19}$$

Eqs. (12) and (19) are the main results of this paper. The heating rate of the second particle is determined from (1), (12) and (19) with obvious replacements

$$\dot{Q}_{12} \to -\dot{Q}_{12} = \dot{Q}_{21},\; \dot{Q}_{13}^{vac} \to \dot{Q}_{23}^{vac},\; \alpha_{1e}''(\omega) \to \alpha_{2e}''(\omega),\; \alpha_{1m}''(\omega) \to \alpha_{2m}''(\omega),\; T_1 \to T_2 \tag{20}$$

Practically, an important characteristic of radiative heat transfer represents the radiative conductance $G$, defined as

$$G = \lim_{T_1 \to T_2} \dot{Q}(T_1, T_2)/|T_1 - T_2| \tag{21}$$

In order to obtain the corresponding functions $G(T)$ at a given temperature $T$, one needs to expand temperature factors in the square brackets of Eqs.(12) and (19) with respect to $T$, assuming $T_1 = T + \Delta T, T_2 = T, T_3 = T$ inserting (12), (19) into (21).

To go a step further beyond the point dipole approximation, we have used an additive summation method to calculate the heat flux between two large spheres at a small gap of width $d$. Introducing $R = 2a + d$ into Eq. (12) (with $a$ being the sphere radius) and integrating over the volume of the spheres $C_1, C_2$ according to the relation

$$1/R^{2n} \rightarrow \iint_{C1C2} (1/R_{1,2}^{2n}) dV_1 dV_2 \ , \ k = 1,2,3, \tag{23}$$

we finally obtain ($x = 2a/(2a+d)$)

$$\dot{Q}_{12} = \frac{2\hbar}{\pi} \int_0^\infty d\omega \omega \left( \text{Im} \left[ \frac{\varepsilon(\omega) - 1}{\varepsilon(\omega) + 2} \right] \right)^2 \left[ 3f_1(x)_1 + (\omega a/c)^2 f_2(x) + (\omega a/c)^4 f_2(x) \right] \cdot \\ \cdot \left[ \coth(\hbar\omega/2k_B T_2) - \coth(\hbar\omega/2k_B T_1) \right] \tag{24}$$

$$f_1(x) = \frac{3}{32} \left[ \ln(1-x^2) + \frac{x^2(2-x^2)}{2(1-x^2)} \right] \tag{25}$$

$$f_2(x) = \frac{3}{32} \left[ 2x^2 - (3x^2 - 2)\ln(1-x^2) - x^3 \ln\frac{1+x}{1-x} \right] \tag{26}$$

$$f_3(x) = \frac{3}{320} \left[ 11x^4 + 2x^2 - (10x^3 - 2x^5) \ln\frac{1+x}{1-x} - (10x^2 - 2)\ln(1-x^2) \right] \tag{27}$$

It is worth noting that in the above derivation of Eq. (24) we have used the expression $\alpha_{1,2}(\omega) = \frac{3}{4\pi} dV_{1,2} \frac{\varepsilon(\omega) - 1}{\varepsilon(\omega) + 2}$ where $\varepsilon(\omega)$ is the frequency–dependent dielectric function of the sphere's material.

## 3. Numerical results

Similar to Ref. [3], we will present our numerical results of the radiative conductance between two silica spheres of equal radii $R_{1,2} = a$ at $T = 300 K$. To calculate the contribution $\dot{Q}_{13}^{vac}$ we assume that vacuum background has the same temperature. The dielectric function of silica is taken from Ref. [17]. The largest contribution to the heat exchange comes from the frequency ranges of 0.04 to 0.07 eV and 0.14 to 0.16 eV. In Figs.2-4 the integral radiative conductance corresponding to one of the spheres is plotted as a function of gap width $d$ at a different radius $a$. In Figs.2,3, lines 1 to 4 correspond to a simple dipole approximation, Eq.(12), referred to as DA, an additive dipole approximation, Eq.(24) (referred to as DAA), the sum of DAA and vacuum contribution, and vacuum contribution alone, respectively. The classical blackbody limit is shown by line 5. In Fig.4 lines 1 and 2 correspond to DA and DAA. The vacuum conductance (19) and classical blackbody limit (26) are shown by lines 3, 4. In the last case we have used the formula

$$G_{BB} = \frac{4\pi^3}{15}\left(\frac{ak_BT}{\hbar c}\right)^2\left(\frac{k_BT}{\hbar}\right)k_B \qquad (26)$$

We see from Figs.2,3 that for small spheres (at $a = 10nm$ and $a = 1\mu m$) the vacuum conductance (lines 4) is small compared to the classical limit, but goes to this one with increasing radius (compare lines 4,5 in Figs.2,3). For spheres with radius of $25\mu m$ the dipole approximation to the vacuum heat conductance becomes incorrect and comparison of Eqs.(19), (26) (lines 3,4 in Fig.4) seems to be of less confidence.

The most intriguing feature of our results as compared to those reported in Ref. [3], obtained at same conditions (see Figs.11,12 in [3]), is the essentially higher value of the radiative conductance. For instance, from our results shown in Figs.2-4 it follows that at a gap of $d = 0.2\,\mu m$ the heat conductance for spheres with radius $10nm, 1\mu m, 25\mu m$ is equal to

$$1.7\cdot 10^{-15}(W/K),\ 5.4\cdot 10^{-10}(W/K),\ 3.3\cdot 10^{-5}(W/K),$$

respectively. On the other hand, the corresponding data from Ref. [3] turn out to be $2\cdot 10^{-16}(W/K), 3\cdot 10^{-11}(W/K)$ and $3.5\cdot 10^{-9}(W/K)$. We see that only the values of the heat conductance at $a = 20\,nm, 1\mu m$ are in reasonable concord (the data in [3] being yet lesser). Also, our data points manifest another distance behavior. What happens is that the corresponding dependences $G(d)$ in [3] show the lesser slope when the gap decreases, with finite value at $d = 0$. This feature is apparent even for a sphere of radius $a = 20nm$. On the contrary, from Figs.2, 3 it follows that the slope of the curves $G(d)$ becomes lesser with increasing the gap width (lines 1 to 3). The slope of line 2 in Fig.4 corresponding to spheres with radius of $25\,\mu m$, proves to be closer to that one obtained in [3] (viz. -0.2 in our work and -0.41 in [3]), but the numerical values of conductance in Fig.4 are larger by 4 order of magnitude.

Such a large discrepancy between our results and those in Ref. [3] for big spheres, in our opinion, unlikely can be attributed to an approximate character of an additive summation method, which turns out to give correct distance dependence in calculations of the Casimir forces and radiative heat exchange in similar configurations [4,13]. As far as concerned to the method developed in [3], the authors have drawn attention to a problem of numerical convergence of the obtained very cumbersome formula for the radiative heat conductance. Evidently, the discussed problem needs further elaboration despite that theoretical predictions made in [3] proved to be in a quite reasonable agreement with the accompanying experimental measurements [6,7].

## 4. Summary

Using a dipole approximation of the fluctuation electromagnetic theory, we have calculated radiative heat flux and conductance in a system of two spherical particles embedded in equilibrium photon gas (vacuum background). The temperatures $T_1, T_2$ of the particles and $T_3$ of vacuum background are assumed to be different. To consider the case when the gap width is smaller than the particle radius, we have used an additive summation method. The obtained formulae take into account both contributions to the heat exchange between the particles and between the particles and vacuum background. We have got significantly larger numerical coefficient in the expression for the dipole rate of the heat exchange between the spherical particles in comparison with [10] and [11,12] ( by $8\pi$ and $16\pi^2$ times). We have also compared our results with recent numerical calculations [3]. At a gap of $0.2\,\mu m$ our results are about 10 times larger for particles with radius of $20 nm$ and $1\mu m$, and by 4 orders of magnitude than for the particles with radius of $25\,\mu m$. Moreover, we have got a quite different distance dependence of the heat conductance : our data points show an increasing slope of the heat conductance dependence at a lesser gap width, and vice versa in Ref. [3]. We believe that such a large discrepancy between our calculations and the theory [3] necessitates further investigation despite the authors have notified a good agreement of the theory [3] with experiment [6,7].

FIGURE CAPTIONS

Fig.1 Two sphere configuration.

Fig.2 Total heat conductance plotted as a function of gap. Both particles have radius of 10 *nm*. Line 1 shows DA, line 2 – DAA, line 3 – the sum of (24) and (19), line 4 – Eq.(19), line 5 – classical black body limit, Eq.(26).

Fig.3 The same as on Fig.2 for particles with radius of $1 \mu m$.

Fig.4 Total heat conductance as a function of gap for big particles with radius of $25 \mu m$. Line 1 –DA, line 2 –DAA, line 3 –Eq.(19), line 4 –Eq.(26).

FIGURE 1

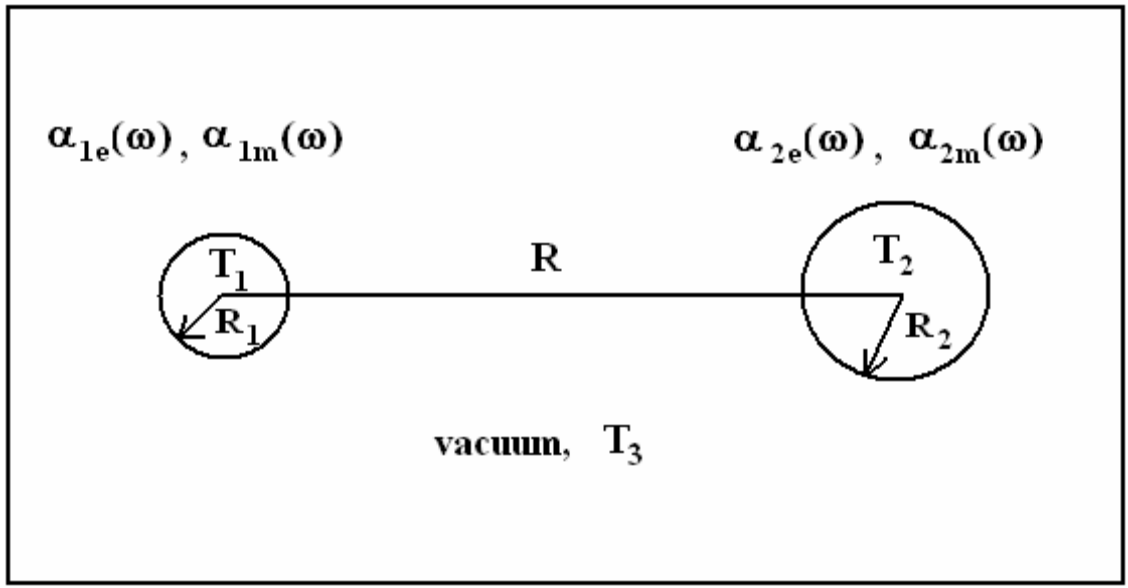

FIGURE 2

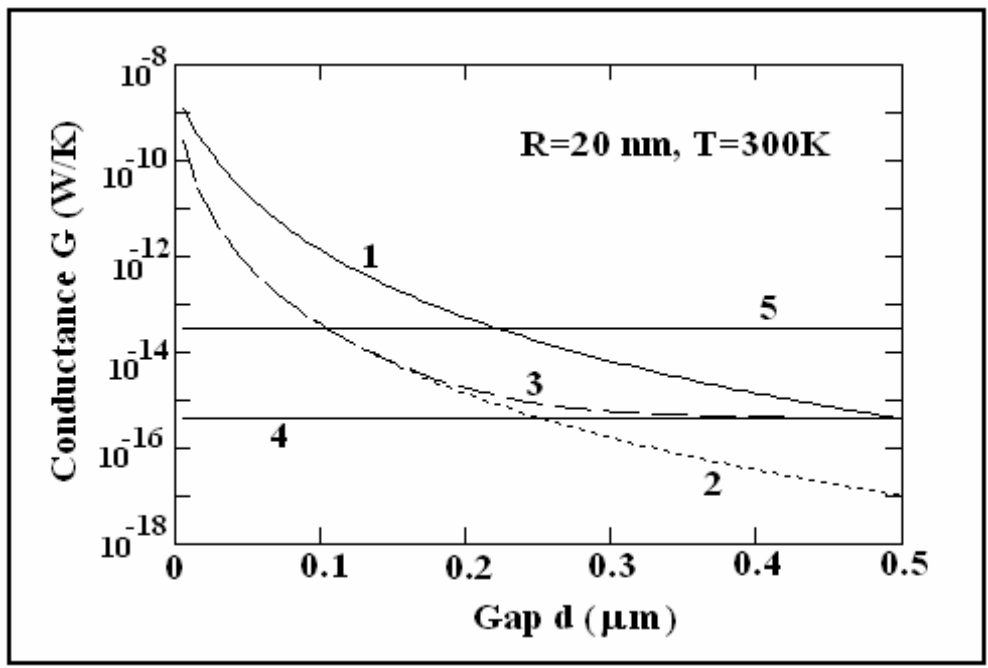

FIGURE 3

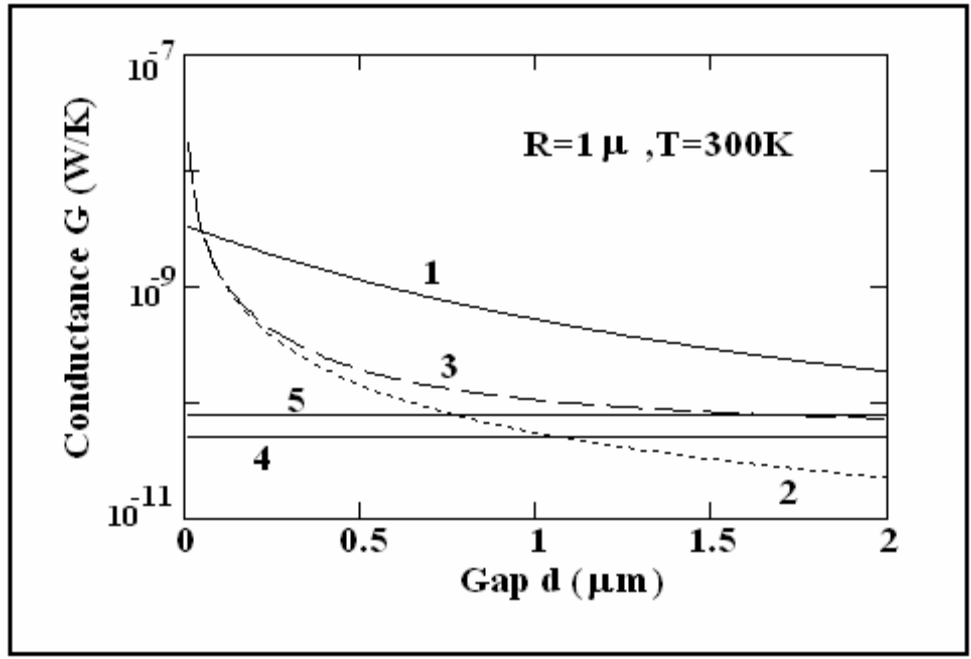

FIGURE 4

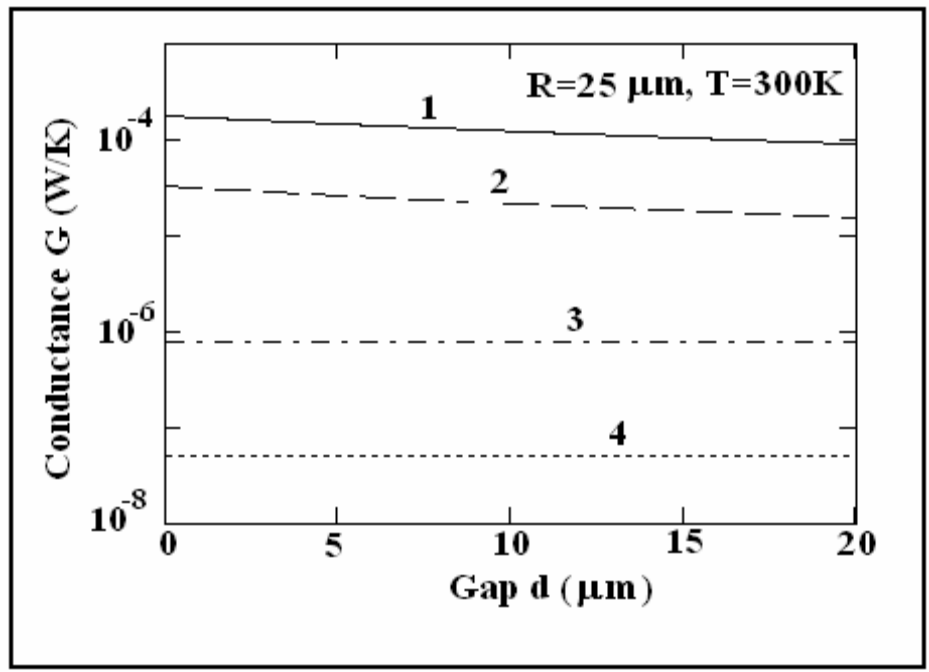